\PassOptionsToPackage{nofiglist}{hypcap}
\documentclass[aps, pra, reprint, superscriptaddress, longbibliography]{revtex4-2}

\usepackage{graphicx}
\usepackage[utf8]{inputenc}
\usepackage[export]{adjustbox}
\usepackage{tikz}
\usepackage{amsmath}
\usepackage{amssymb}
\usepackage{mathtools}
\usepackage{hyperref}
\usepackage{graphicx}
\usepackage{epstopdf}
\usepackage{dcolumn}
\usepackage{soul}
\usepackage{mathrsfs}
\usepackage{bm}
\usepackage{soul}
\usepackage{xspace}
\usepackage{verbatim}
\usepackage{color}
\usepackage{color,soul}
\usepackage[normalem]{ulem}
\usepackage{xcolor}

\newcommand{\ket}[1]{\left| #1 \right\rangle}

\hypersetup{colorlinks=true,linkcolor=blue,citecolor=blue, filecolor=blue,urlcolor=blue,breaklinks=true}

\definecolor{lime}{HTML}{A6CE39}
\DeclareRobustCommand{\orcidicon}{%
	\begin{tikzpicture}
	\draw[lime, fill=lime] (0,0) 
	circle [radius=0.16] 
	node[white] {{\fontfamily{qag}\selectfont \tiny ID}};
	\draw[white, fill=white] (-0.0625,0.095) 
	circle [radius=0.007];
	\end{tikzpicture}
	\hspace{-2mm}
}

\foreach \x in {A, ..., Z}{%
	\expandafter\xdef\csname orcid\x\endcsname{\noexpand\href{https://orcid.org/\csname orcidauthor\x\endcsname}{\noexpand\orcidicon}}
}

\begin{document} 

\title[F]
{Floquet-engineered system-reservoir interaction in the transverse field Ising model}

\author{Maritza Ahumada \orcidA{}}
\email{maritza.ahumada@usm.cl}
\affiliation{Departamento de F\'isica, Universidad de Santiago de Chile (USACH), Avenida V\'ictor Jara 3493, 9170124, Santiago, Chile}

\author{Natalia Valderrama-Quinteros\orcidB{}}
\affiliation{Departamento de F\'isica, Universidad de Santiago de Chile (USACH), Avenida V\'ictor Jara 3493, 9170124, Santiago, Chile}



\author{Guillermo Romero \orcidC{}}
\affiliation{Departamento de F\'isica, CEDENNA, Universidad de Santiago de Chile (USACH), Avenida V\'ictor Jara 3493, 9170124, Santiago, Chile}

\begin{abstract}
Periodically driving a quantum many-body system can drastically change its properties, leading to exotic non-equilibrium states of matter without a static analog. In this scenario, parametric resonances and the complexity of an interacting many-body system are pivotal in establishing non-equilibrium states. We report on a Floquet-engineered transverse field Ising model for the controlled propagation in one dimension of spin waves. The underlying mechanisms behind our proposal rely on high-frequency drivings using characteristic parametric resonances of the spin lattice. Many-body resonances modulating spin-sping exchange or individual spin gaps inhibit interactions between spins thus proving a mechanism for controlling spin-wave propagation and a quantum switch. Our schemes may be implemented in circuit QED with direct applications in coupling-decoupling schemes for system-reservoir interaction and routing in quantum networks.     
\end{abstract}

\maketitle



\section{Introduction}
A common requirement for implementing various quantum technologies---such as quantum batteries~\cite{Campaioli2024Rev,Araya2024}, quantum thermal engines~\cite{Cangemi2024,centamori2023}, quantum state transfer devices~\cite{Yin2013,Li2018,HuiZhou_2019}, and single-photon coherent transport~\cite{Zhou2009,Liao2009}---is the precise control of coupling and decoupling in many-body systems.
For instance, a key component in the thermodynamic cycles of a quantum battery is the ability to precisely connect and disconnect the battery from the charger, requiring fine control over their coupling. Recently, Araya et al. \cite{Araya2024} introduced a novel cycle incorporating a memory element. The protocol was tested using a one-dimensional (1D) transverse spin-1/2 Ising chain, where one spin serves as the battery while the remaining spins act as the charger. Moreover, in the internal processes of a thermodynamic cycle in a quantum thermal engine, a fundamental requirement is the ability to connect and disconnect the working system from external reservoirs. In this context, Centamori et al. \cite{centamori2023} investigated a quantum engine based on a 1D spin-1/2 chain. Also, in quantum state transfer using spin-1/2 chains for applications in quantum communication and computing, precise control over the release of states from a node to the quantum channel is crucial for an effective distribution across a quantum network ~\cite{HuiZhou_2019}.

Floquet engineering~\cite{Floquet1883,Grifoni1998Oct,Bukov2015Mar,Oka2019,eisert2015} provides a framework for implementing various applications such as Floquet sensing~\cite{Mishra2021Aug,Mishra2022Aug}, and suits perfectly for coupling-decoupling mechanisms in quantum many-body systems~\cite{Grossmann1991Jul,Kayanuma2008Jan,Gong2009Sep}. In a recent contribution, Peña et al.~\cite{Pena2024Jan} proposed controlling the nonequilibrium many-body dynamics of a one-dimensional spin-1/2 lattice with periodically modulated bonds, leveraging parametric resonances. This approach produced spatial and temporal localization of correlated spin pairs via dynamically breaking correlated spin pairs from the edges towards the center of the lattice. The creation of correlated spin pair fluctuations also implies a spin-blocking effect acting on a three-spin lattice. The latter presents itself as a viable mechanism for coupling-decoupling system-reservoir interaction and controlling the dynamics of a many-body system.

This work addresses coupling and decoupling requirements in many-body systems, and provides a versatile and realistic scheme for implementing an on-off switch to control the propagation of pulses in a 1D-dimensional Ising spin-1/2 lattices, specifically tailored for superconducting circuit platforms.
We present a Floquet-engineered transverse field Ising model that enables subsystems' controlled coupling, decoupling, and isolation in spin-1/2 lattices. Our approach relies on high-frequency driving, which leverages characteristic parametric resonances of the spin-lattice~\cite{Pena2024Jan}. We abord the mechanisms where these driving frequencies can act on either the spin-spin exchange or individual spins. The resulting mechanisms correspond to a quantum switch, which is analyzed through the dynamics of spin and correlation waves in a 1D lattice. Our findings may have applications in coupling-decoupling schemes for system-reservoir interaction, and routing in quantum networks.      


\section{The Model}
\label{ModulBond}
\begin{figure}[h]
\centering
\includegraphics[width=\columnwidth]{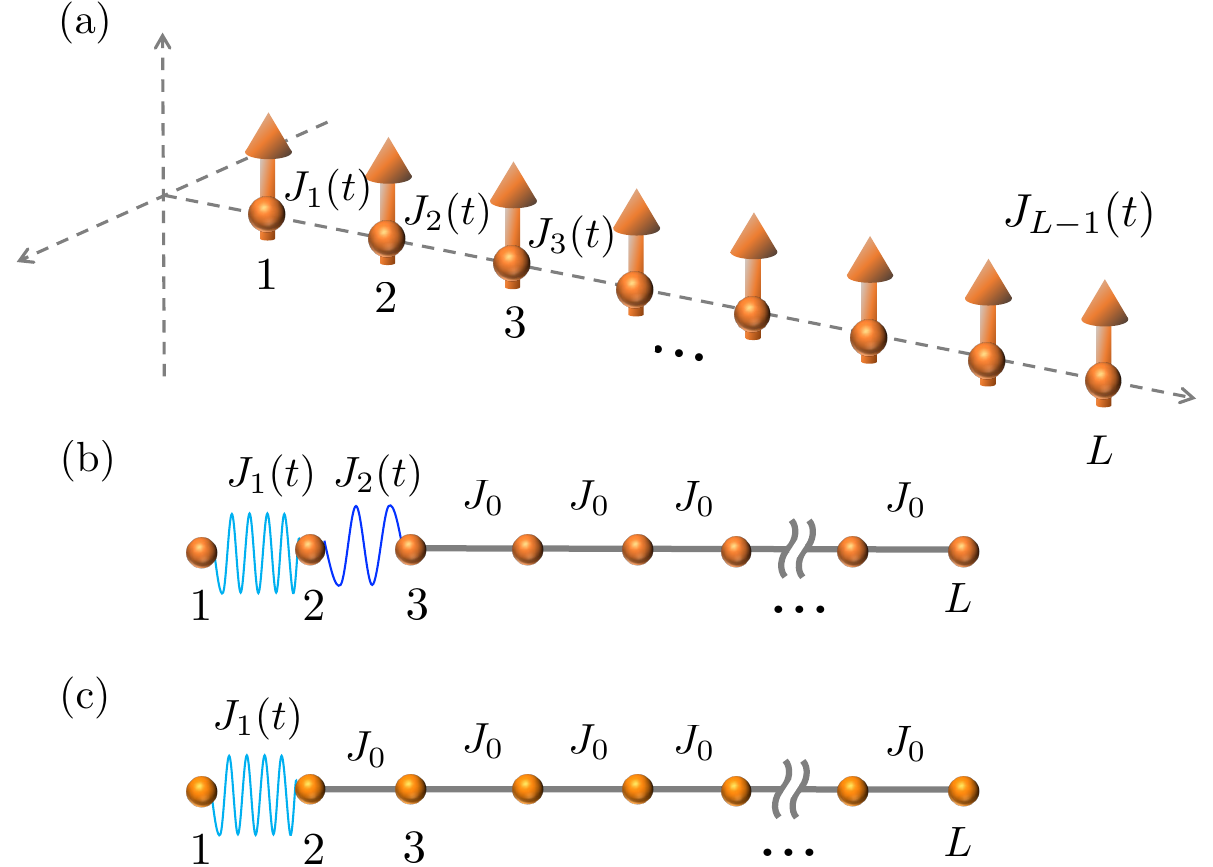}
\caption{
Schematic view of the 1D spin-1/2 lattice: (a) View of the system configuration at $t=0$, where the initial condition is $\ket{\psi_0}=\bigotimes_{j=1}^L\ket{\uparrow_j}$. The spin-exchange magnitude $J_{\alpha}(t)$, with  $\alpha \in [1,L-1]$ the bond index, is a time-dependent function coupling nearest-neighbor spins. (b) Configuration with the first periodic modulated bond given by $J_{1}(t)=J_{0}\cos{(\Omega_{1}t)}$, represented by the light blue wavy line, and the second bond $J_{2}(t)=J_{0}\cos{(\Omega_{2}t)}$ by the blue wavy line. (c) Configuration with bonds modulated as $J_{1}(t)=J_{0}\cos{(\Omega_{1}t)}$ and $J_{2}=J_{0}$. 
}
\label{Fig1}
\end{figure}

The proposed system comprises a 1D spin-1/2 lattice with $L$ sites, nearest-neighbor interactions, and periodically modulated bonds as is depicted schematically in Fig.~\ref{Fig1}(a).
We consider a time-dependent transverse-field Ising model with open boundary conditions \cite{Hong2022Aug}, where the Hamiltonian $H(t)$ reads 
\begin{equation}
H(t)= \hbar g \,\sum_{j=1}^{L}\hat{\sigma}_{j}^{z}+\hbar\,\sum_{j=1}^{L-1}J_{j}(t)\,\hat{\sigma}_{j}^{x}\hat{\sigma}_{j+1}^{x}. 
\label{H1}
\end{equation}
Here, $\hat{\sigma}^{x,z}_j$ denotes the Pauli matrices at the $j$th lattice site, and $g$ represents the transverse magnetic field. The spin-exchange magnitude $J_{\alpha}(t)=J_{0}\cos(\Omega_{\alpha} t)$, with  $\alpha \in [1,L-1]$ the bond index, is a time-dependent function coupling nearest-neighbor spins, where $J_{0}$ is the bare exchange coupling, and $\Omega_{\alpha}$ is the driving frequency.

One of the key elements of the system is the three leftmost spins in the lattice. Here, the first bond is driven with a frequency $\Omega_{1}=4g$, whereas the second bond is driven with a frequency $\Omega_{2}=2g$, as schematized in Fig.~\ref{Fig1}(b).
We choose these driving frequencies such that the three spins, initialized in the product state $\ket{\psi_0}=\ket{\uparrow\uparrow\uparrow}$, exhibit correlated spin pairs~\cite{Pena2024Jan}. Thus, the spin at site $j=3$ experiences a blocking effect, due to high frequencies modulating the spin exchange,  with negligible fluctuations around the average $\langle \hat{\sigma}^z_3\rangle=1$. In contrast, the spins located at the $j=1$ and $j=2$ sites oscillate between separable and quasi-maximally entangled states following a subharmonic response in their local observables~\cite{Pena2024Jan}, thus breaking the discrete-time translational symmetry \cite{Sacha2015Mar,Else2016Aug,Yao2017Jan,Sacha2017Nov,Pizzi2019Oct,TimeCrystals,Else2020Mar,Pizzi2021Apr,Pizzi2021Sep0,Pizzi2021Sep1,Ye2021Sep,Ippoliti2021Sep}. The dynamics of the third spin is unblocked by turning off the driving frequency, namely, $\Omega_{2}=0$, which reduces the second modulate bond to a constant spin-spin exchange $J_{2}=J_{0}$. The last case is shown in Fig.~\ref{Fig1}(c), where a spin wave will propagate along the lattice. 

We explore the cases shown in Fig.\ref{Fig1} where the initial state of the system is $\ket{\psi_0}=\bigotimes_{j=1}^L\ket{\uparrow_j}$, therefore, at $t=0$ all spins have average value $\langle\hat{\sigma}^{z}_{j}\rangle=1$ (c.f. Fig.\ref{Fig1}(a)). We solve the spin-lattice dynamic numerically, computing the state $|\psi(t)\rangle=\hat{U}(t,0)|\psi_{0}\rangle$, where the evolution operator reads
\begin{equation}
 \hat{U}(t,0)=\hat{T}\,\exp\left({\frac{-i}{\hbar}\int_{0}^{t}H(s)\,ds}\right),
\label{EvolOper}
\end{equation}
with $\hat{T}$ denoting the time-ordering operator. 

In the following, we perform numerical analysis using exact diagonalization with the QuSpin~\cite{Weinberg2017, Weinberg2019} library up to $L=16$ spins. Here we use Hamiltonian parameters in terms of the transverse field $g$, that is, $J_0=0.1g$, and time step $\delta t=5\times 10^{-3}~T$, where $T=\pi/g$ is the period of the Hamiltonian (\ref{H1}). The dynamic properties of the lattice are characterized by the average value $\langle \sigma_j^z \rangle$ for each spin, and the nearest-neighbor correlation functions between consecutive pairs of spins 
\begin{equation}
C_{j,j+1}(t)=\langle\,\hat{\sigma}^{z}_{j} \hat{\sigma}^{z}_{j+1}\,\rangle -\langle\,\hat{\sigma}^{z}_{j}\,\rangle\langle\,\hat{\sigma}^{z}_{j+1}\,\rangle.
\label{correlation function}
\end{equation}
The average value $\langle \hat{\sigma}_j^z \rangle$ and two-point correlation functions can be measured in quantum simulators such as trapped ions \cite{Richerme2014Jul} or superconducting circuits \cite{Zhao2022Oct,Song2024Apr,Gong2021Jul}.

\begin{figure}
\centering
\includegraphics[width=\columnwidth]{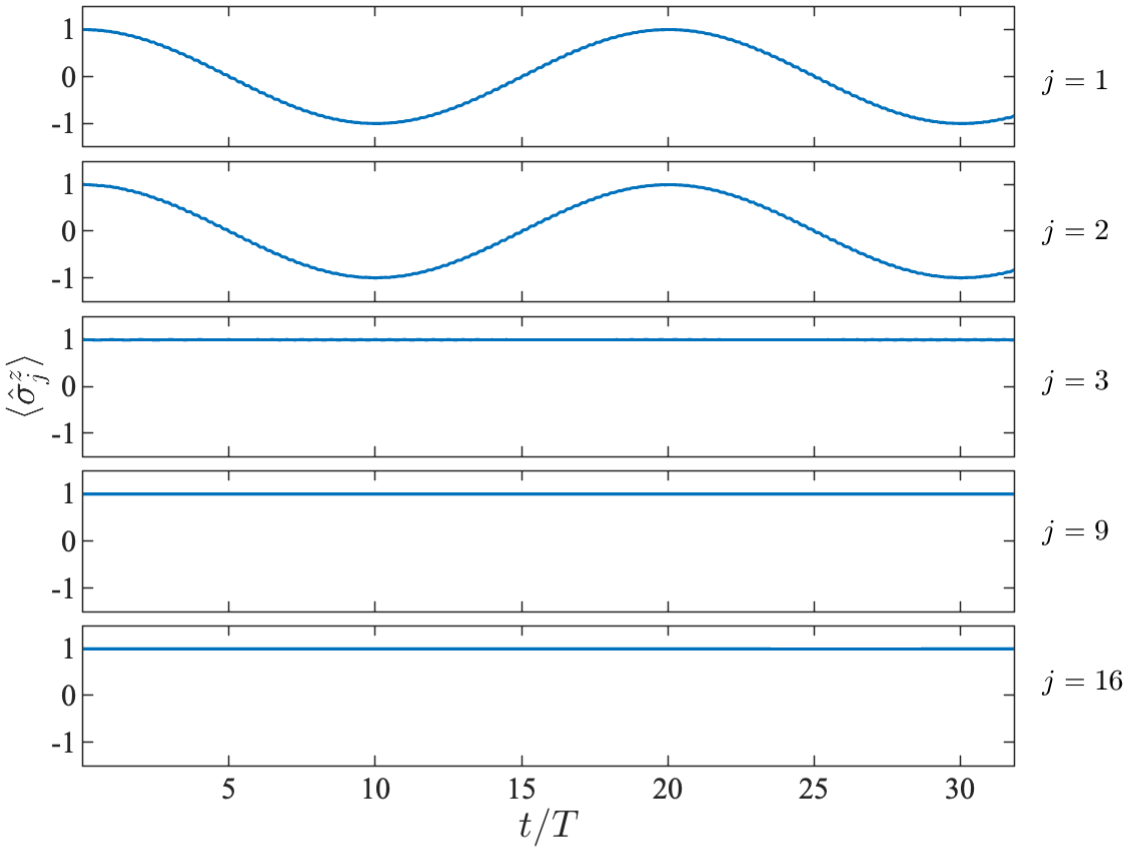}
\caption{Average value $\langle\hat{\sigma}^{z}_{j}\rangle$ at the $j$th spin as a function of time in units of the Hamiltonian period $T$. The parameters are defined in terms of the transverse field $g$, that is, $J_{0}=0.1\,g$, $\Omega_{1}=4\,g$, and $\Omega_{2}=\Omega_{1}/2$. In this simulation, we consider a lattice of $L=16$ spins.} 
\label{Fig2}
\end{figure}

 We start exploring the dynamic properties of the 1D lattice using the configuration of Fig.~\ref{Fig1}(b). Figure~\ref{Fig2} shows the average value $\langle\hat{\sigma}^{z}_{j}\rangle$ as a function of time in units of $T$, for the first three lattice sites  ($j=1,2,3$) and other representative sites of the lattice ($j=9,16$). The average value $\langle\hat{\sigma}^{z}_{j}\rangle$ of the first two spins located at the left side ($j=1,2$) has a subharmonic response with a period  $T'=20T$, as shown in panels $j=1,2$ of Fig.~\ref{Fig2}. Such a behavior corresponds to correlated spin pairs, which induces a spin-blocking effect over the third spin, which experiences negligibly fluctuations around the average $\langle\hat{\sigma}^{z}_{3}\rangle = 1$, as shown in the panel $j=3$ of Fig.~\ref{Fig2}. Such blocking effect inhibits the spreading of spin waves in the 1D lattice, and therefore the fluctuations around the average $\langle\hat{\sigma}^{z}_{j}\rangle$ of all spins at sites $j>3$ are negligible, as shown in panels $j=9,16$ of Fig.~\ref{Fig2}. 

\begin{figure}
\centering
\includegraphics[width=\columnwidth]{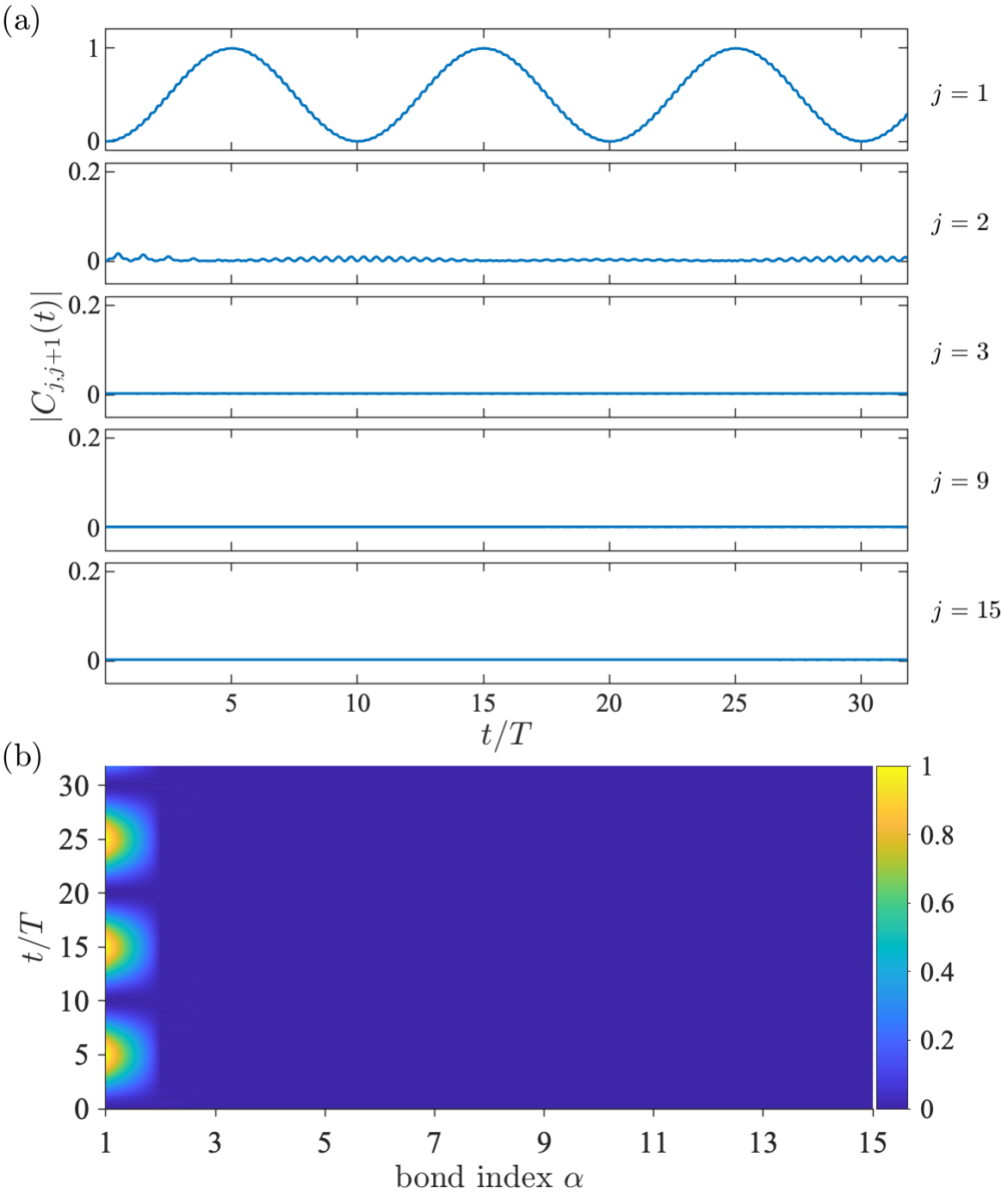}
\caption{(a) Nearest-neighbor correlation $|C_{j,j+1}(t)|$ as a function of time in units of the Hamiltonian period $T$. The panels correspond to correlations functions from $|C_{1,2}(t)|$ to $|C_{3,4}(t)|$, and other representative correlations functions $|C_{9,10}(t)|$ and $|C_{15,16}(t)|$.
(b) $|C_{j,j+1}(t)|$ as a function of time in units of $T$ and the lattice bond index. 
The parameters are defined in terms of the transverse field $g$, that is, $J_{0}=0.1\,g$, $\Omega_{1}=4\,g$, and $\Omega_{2}=\Omega_{1}/2$. In this simulation, we consider a lattice with $L=16$ spins. }
\label{Fig3}
\end{figure}

Moreover, the above configuration induces a blocking effect of the correlation propagation through the lattice. 
Figure~\ref{Fig3} shows the absolute value of nearest-neighbor correlation $|C_{j,j+1}(t)|$ as a function of time. Initially, the correlations of all pairs in the lattice are $C_{j,j+1}(0)=0$.
Figure~\ref{Fig3}(a) shows the oscillatory behavior of the spin-pair correlation $|C_{1,2}(t)|$, which is obtained due to a subharmonic response induced by the parametric resonance. The correlation $|C_{2,3}(t)|$ oscillates around zero due to the blocking effect of the third spin. Thus, the correlation of all consecutive pairs in the lattice reduces to zero, and the correlation wave through the lattice is blocked, as shown in Fig.~\ref{Fig3}(b).

\begin{figure}
\centering
\includegraphics[width=\columnwidth]{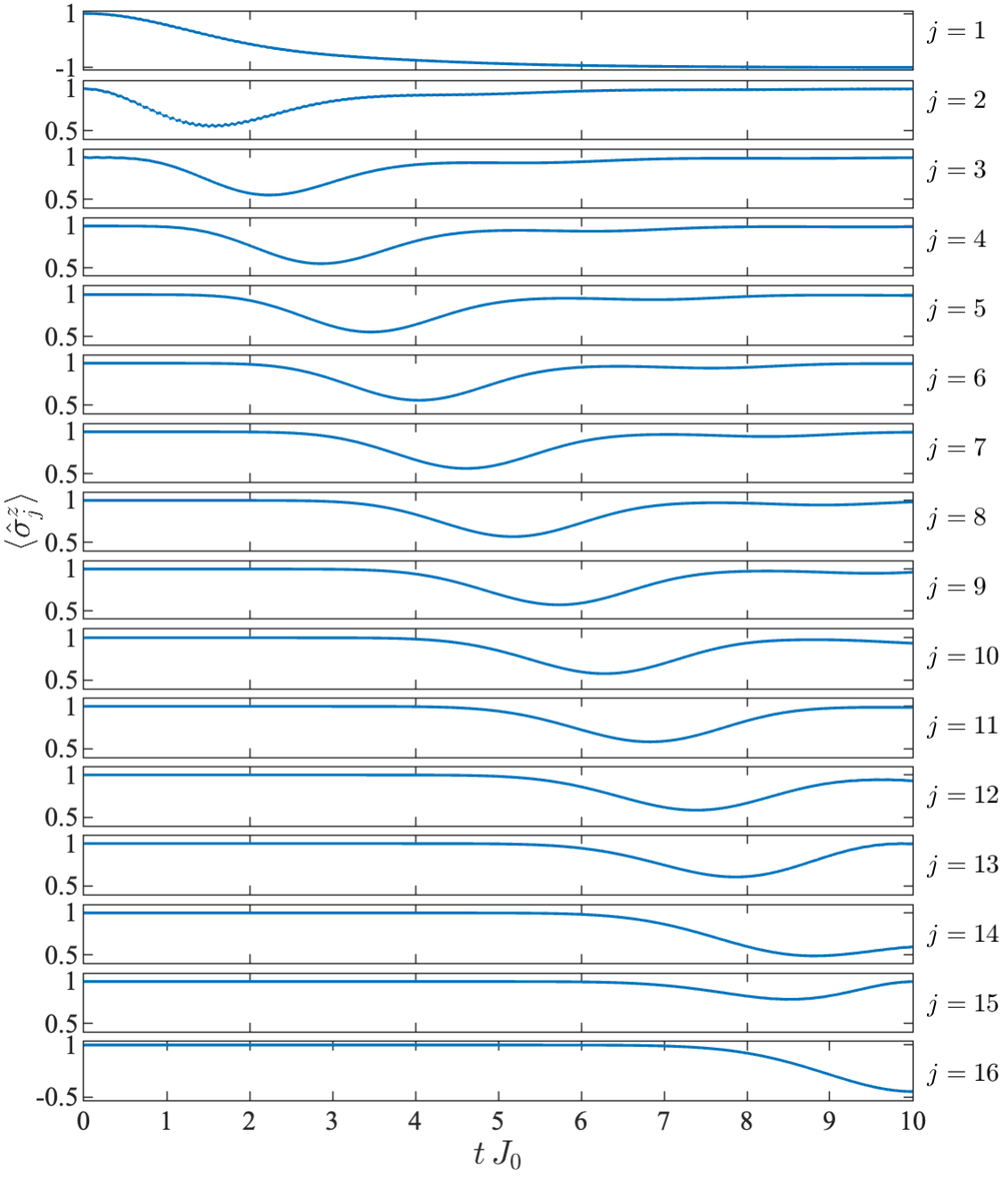}
\caption{Average value $\langle\hat{\sigma}^{z}_{j}\rangle$ at the $j$th spin as a function of time. The panels describe all the spin-lattice sites $1\leq j\leq 16$. The parameters are defined in terms of the transverse field $g$, that is, $J_{0}=0.1\,g$, $\Omega_{1}=4\,g$, and $\Omega_{2}=0$. In this simulation, we consider a lattice of $L=16$ spins.} 
\label{Fig4}
\end{figure}

\begin{figure}
\includegraphics[width=\columnwidth]{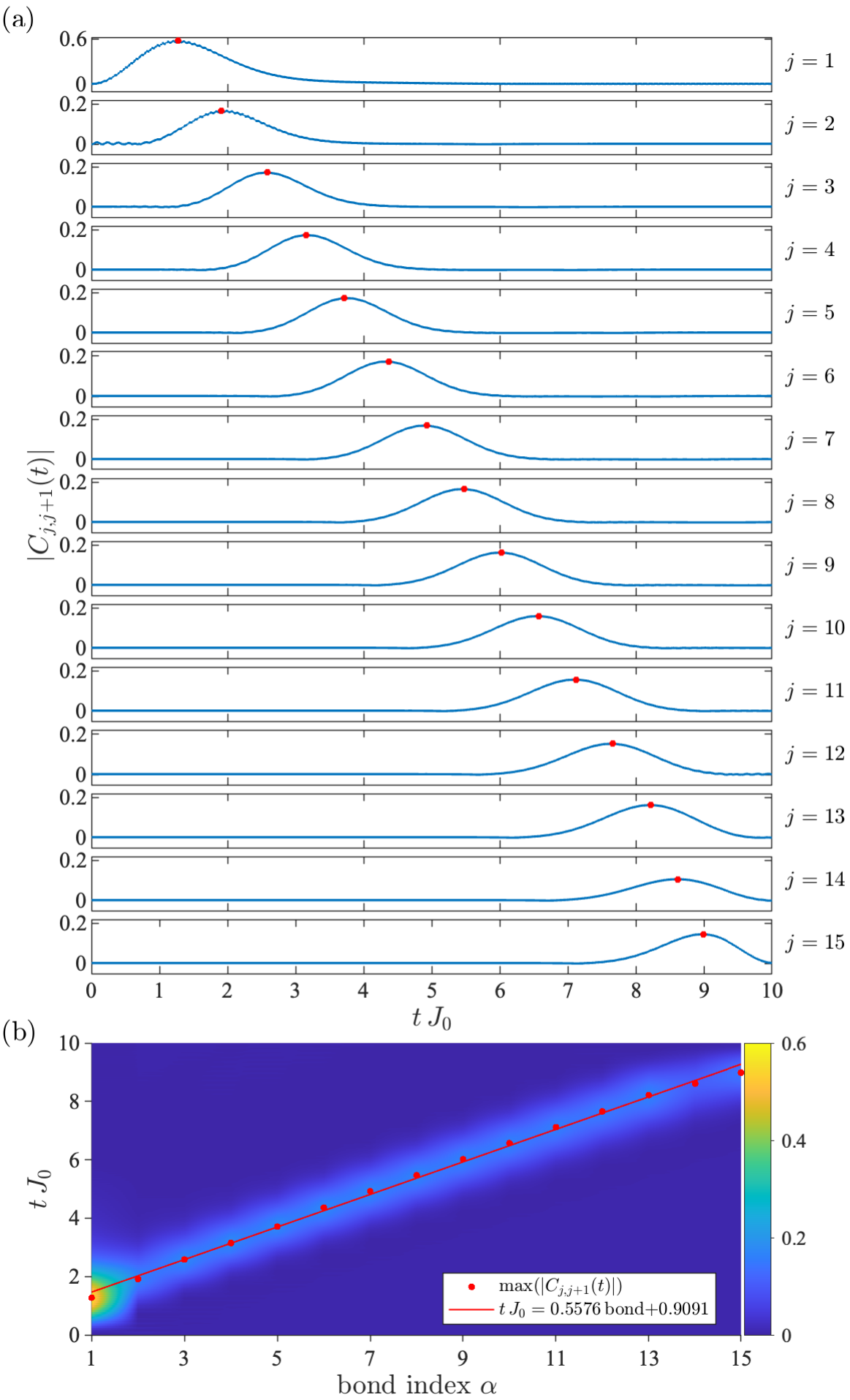}
\caption{(a) Nearest-neighbor correlation functions $|(C_{j,j+1})(t)|$ as a function of time in units of $J_{0}$. The panels correspond to correlations functions from $|C_{1,2}(t)|$ to $|C_{15,16}(t)|$. The red dots indicate the maximum value of $|C_{j,j+1}(t)|$. (b) Contour plot: Two-point correlation functions as a function of time and the lattice bond index $\alpha$. The red solid line corresponds to the linear fitting of $\text{max}(|C_{j,j+1}(t)|)$. The parameters are defined in terms of the transverse field $g$, that is, $J_{0}=0.1\,g$, $\Omega_{1}=4\,g$, and $\Omega_{2}=0$. In this simulation, we consider a lattice of $L=16$ spins.}
\label{Fig5}
\end{figure}

Now, we present the dynamical properties of the 1D lattice in the configuration shown in Fig.~\ref{Fig1}(c). Here, we use Hamiltonian parameters in terms of the transverse field $g$, that is, $J_0=0.1g$, and time step $\delta t J_0=0.00125$.
Figure~\ref{Fig4} shows the average value $\langle\hat{\sigma}^{z}_{j}\rangle$ as a function of time for all lattice sites. The configuration of the three first spins allows the propagation of a spin wave through the lattice due to the unblocking of the third spin, as shown in Fig.~\ref{Fig4}. Notably, the Floquet engineering of the first bond is crucial for generating such spin waves. Otherwise, the spin lattice remains static. 

The behavior of average values $\langle\hat{\sigma}^{z}_{j}\rangle$ shown in Fig.~\ref{Fig4} can be explained as follows. At time $t$, the wave function reads
\begin{equation}
\begin{aligned}
    |\psi(t)\,\rangle =& \,\tilde{\alpha}(t)\,|\uparrow,\uparrow,...\uparrow_{L}\,\rangle \\&+ \sum_{j=2}^{L}\tilde{\beta}_{j}(t)|\downarrow_{1},\uparrow,...,\uparrow,\downarrow_{j},\uparrow,...,\uparrow\,\rangle\,,
\end{aligned}
\end{equation}
where $\tilde{\alpha}(t)$ is the probability amplitude of the initial state, and $\tilde{\beta}_j(t)$ is the amplitude of the state where the first spin is down, and a spin flip occurs in the $j$th site.

The modulated bond between the two first spins $J_{1}(t)=J_{0}\cos{(\Omega_1\,t)}$ allows the oscillation of the spins at $j=1,2$ thus creating a transient situation where the whole spin-lattice may be described by a linear combination of states $|\uparrow,\uparrow,\uparrow,...\uparrow_{L}\,\rangle$ and $|\downarrow,\downarrow,\uparrow,...\uparrow_{L}\,\rangle$. 
Since the second bond is constant $J_{2}=J_{0}$, the spin-flip exchange is possible between the second and third spins, and the first spin keeps the down state. Such spin flip exchange propagates throughout the lattice, allowing a spin wave propagation [see panels $j>2$ of Fig.~\ref{Fig4}].

Figure~\ref{Fig5}(a) shows the absolute value of nearest-neighbor correlation functions $|C_{j,j+1}(t)|$ as a function of time. Figure~\ref{Fig5} clearly shows a correlation wave's propagation through the lattice. To give an estimation of the group velocity of such correlation wave, we identify the maximum value of the correlation functions for each spin pair, which are indicated with red dots in Figs.~\ref{Fig5}(a) and~\ref{Fig5}(b). The red solid line in Fig.~\ref{Fig5}(b) indicates the linear fitting of the correlation maxima, which follows the linear relation $tJ_{0}=0.5576\,\text{bond}+0.9091$, where the group velocity is $v^{\text{group}}=1.7934\,J_{0}$. The latter is under the upper limit given by the Lieb-Robinson velocity \cite{lieb-robinson1972} in 1D transverse-field Ising models, $v_{\rm LR}=2\,J_{0}$ \cite{bravyi2006jul,cheneau2012light,gong2022,Kaneko2023}.

\section{Quantum switch}

\begin{figure}
\includegraphics[width=\columnwidth]{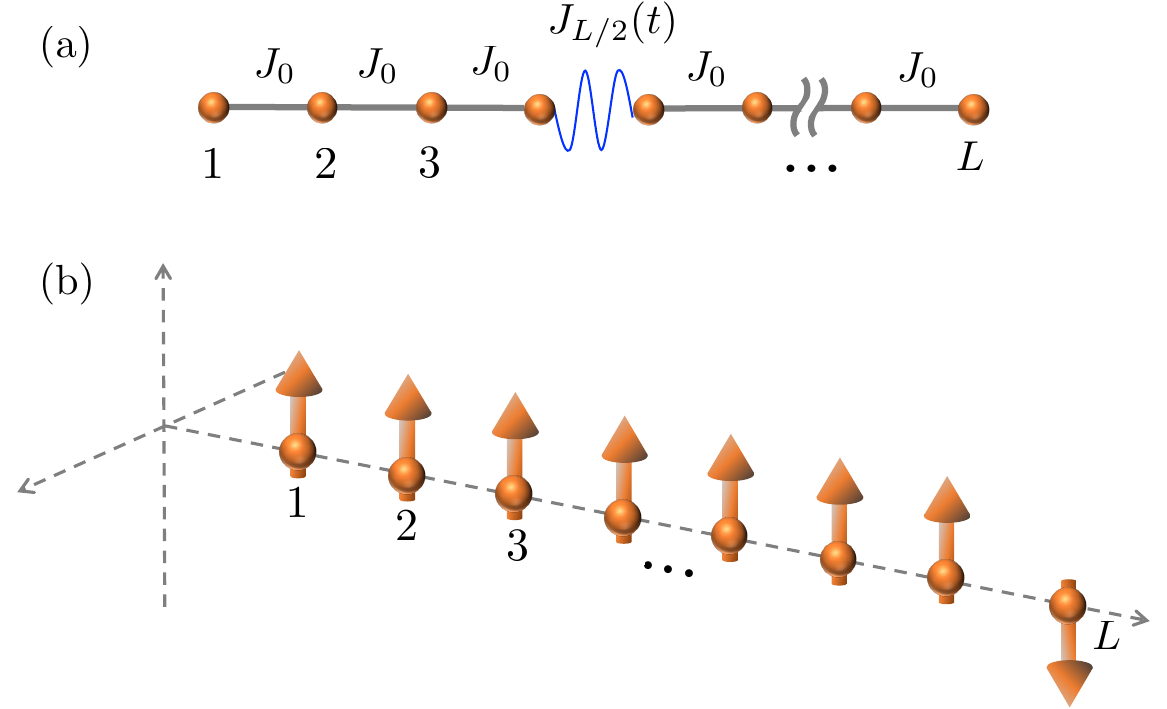}
\caption{(a) Configuration with the $L/2$-th bond modulated by $J_{L/2}(t)=J_{0}\cos{(\Omega\,t)}$, considering $L$ even. The blue wavy line represents the modulated bond. The rest of the lattice bonds are given by constant spin-spin exchange, $J_{\alpha}=J_{0}$. (b) Schematic of the system configuration at $t=0$. The initial condition is given by $\ket{\psi_0}=\bigotimes_{j=1}^{L-1}\ket{\uparrow_j}\otimes\ket{\downarrow_L}$.} 
\label{Fig6}
\end{figure}

In this section, we present the dynamics of the 1D spin-1/2 lattice using a single modulated bond as a switch of propagating spin waves. We choose the $L/2$-th bond as $J_{L/2}(t)=J_{0}\cos{(\Omega\,t)}$, where the driving frequency $\Omega$ is the switch control parameter. The rest of the bonds are constant, $J_{0}$, as shown schematically in Fig.~\ref{Fig6}(a). To characterize the operational mode of the lattice, i.e., blocking (switch off) or unblocking (switch on) mode of the spin wave, we analyze the correlation function between the sites $L/2$ and $L/2+1$ for frequency values in the range $0\leq \Omega/g \leq 3$. We explore the case where initially all spins have average value $\langle\hat{\sigma}^{z}_{j}\rangle=1$ but the last spin at the site $j=L$ has $\langle\hat{\sigma}^{z}_{j}\rangle=-1$; see Fig.~\ref{Fig6}(b). Therefore, the initial state reads $\ket{\psi_0}=\bigotimes_{j=1}^{L-1}\ket{\uparrow_j}\otimes\ket{\downarrow_L}$, and the spin wave propagation occurs initially from right to left in the lattice.

\begin{figure}
\includegraphics[width=\columnwidth]{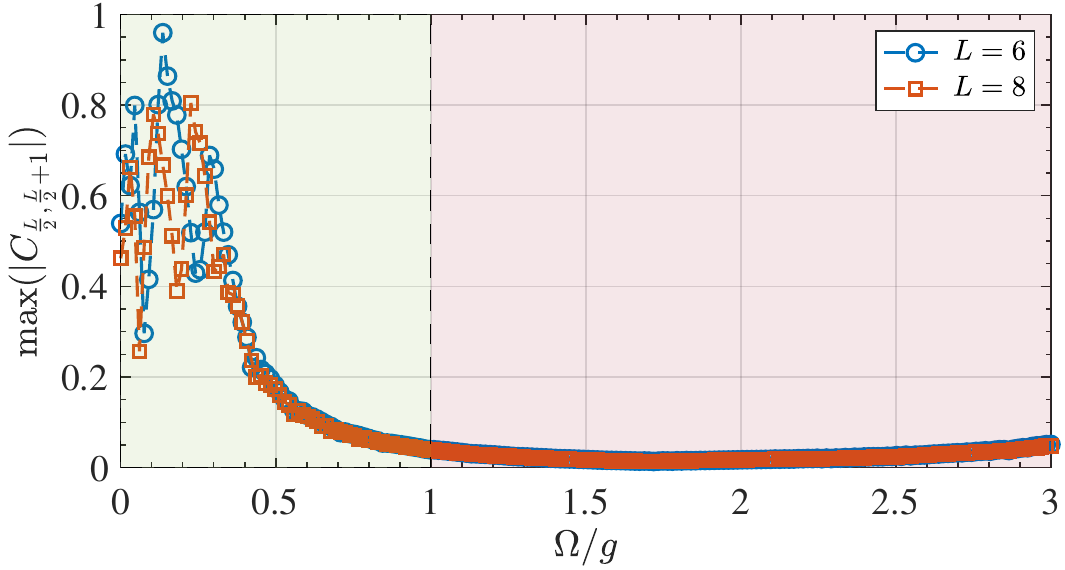}
\caption{The operational mode of the lattice: The maximum of the absolute value of correlation function ${\rm max}|C_{L/2,L/2+1}(t)|$ as a function of $\Omega$ in units of $g$ for $L=6$ (blue circles) and $L=8$ sites (orange squares).
The green area indicates the values of $\Omega$ where the lattice is in the spin wave's unblocking mode (switch on), and the red area indicates the blocking mode (switch off). The parameters are defined in units of the transverse field $g$, that is, $J_{0}=0.1\,g$. The final time of simulation for each curve $C_{L/2,L/2+1}(t)$ is $t_{f}=100\,J_{0}^{-1}$. The driving frequency $\Omega$ $\in\,[0.0\,g,3.0\,g]$ with spacing of $\delta\Omega=0.0151\,g$.} 
\label{Fig7}
\end{figure}

Using exact diagonalization, we numerically solved the Schr\"odinger equation for small lattices of $L=6$ and $L=8$ sites and for several values of the driving frequency $\Omega$ up to the final time $t_{f}=100J_0^{-1}$ which is enough to capture nontrivial dynamics along the small lattices. Since the correlation function $C_{L/2,L/2+1}(t)$ (see Eq.~(\ref{correlation function})) exhibits an oscillatory behavior, we consider the maximum of its absolute value, ${\rm max}|C_{L/2,L/2+1}|$, as a representative quantity of created correlations. Figure~\ref{Fig7} shows the results where we identify the green area as the spin wave's unblocking mode (switch on), and the red area indicates the blocking mode (switch off). The latter is characterized by ${\rm max}|C_{L/2,L/2+1}|\approx 0.05$. Both regions have a broad range of frequencies for performing the two operational modes of the lattice.  

\begin{figure}[h!]
\includegraphics[width=\columnwidth]{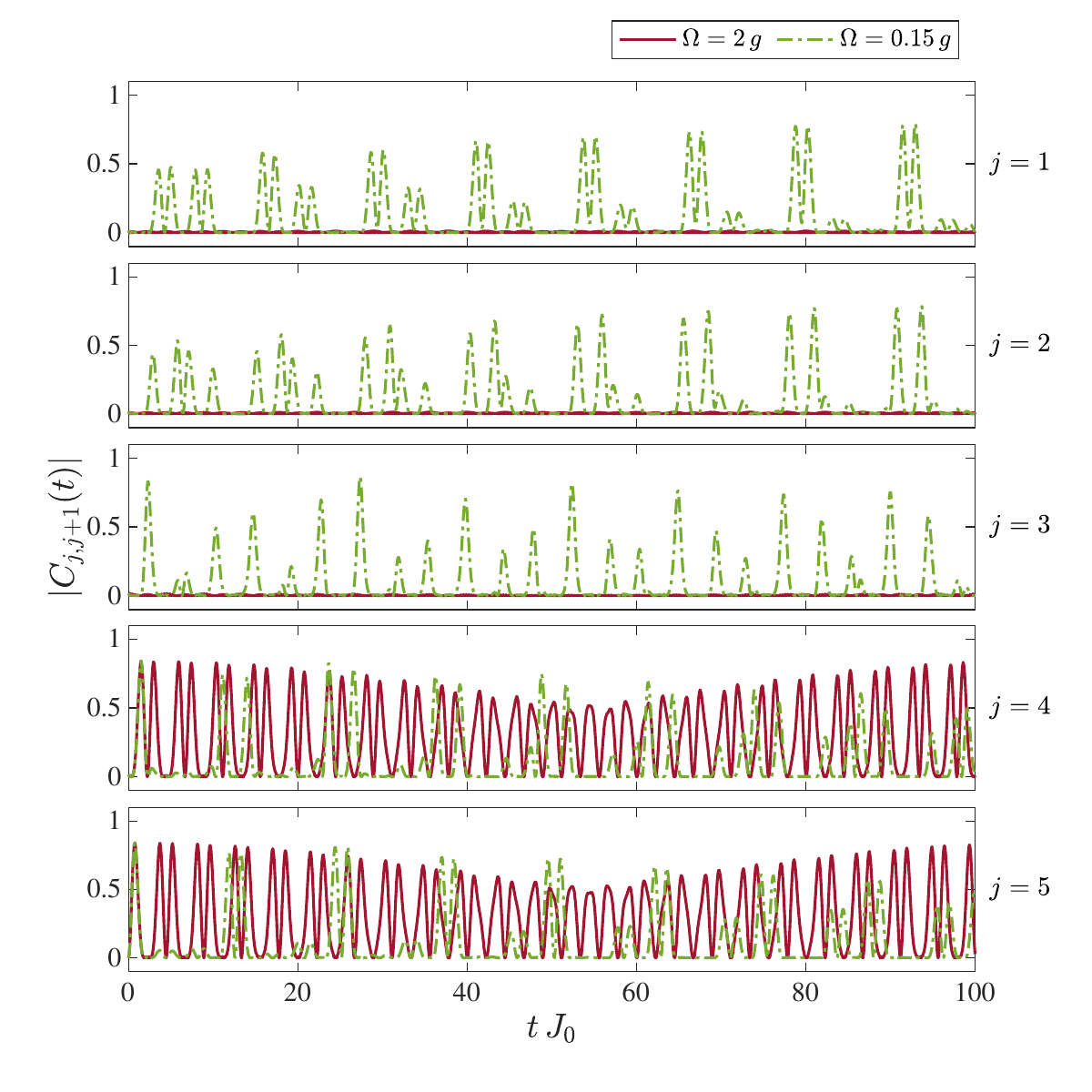}
\caption{Absolute value of correlation function $|C_{j,j+1}(t)|$ as a function of time, for switch on with $\Omega=0.15\,g$ (green dashed line) and switch off with $\Omega=2\,g$ (red solid line). The panels correspond to correlations function from $|C_{1,2}(t)|$ to $|C_{5,6}(t)|$. This simulation considers a lattice of $L=6$ spins and $J_{0}=0.1\,g$.} 
\label{Fig8}
\end{figure}

Figure~\ref{Fig8} shows the correlation functions $|C_{j,j+1}(t)|$ as a function of time for a lattice of $L=6$ spins. Each panel shows the correlation function for two values of the driving frequency. Red lines stand for $\Omega=2\,g$ within the switch-off region, and green-dashed lines stand for $\Omega=0.15\,g$ within the switch-on region (c.f. Fig.~\ref{Fig7}). When the driving frequency is chosen within the switch-off region, the left side of the lattice remains uncoupled from the right side. This is evidenced by the correlation functions of the lattice left side, which remain nearly zero into the panels $j\leq L/2$ of Fig.~\ref{Fig8}. Nontrivial dynamics of $|C_{j,j+1}(t)|$ occurs only on the right side of the lattice ($j>L/2$), and it is blocked toward the left side. 
The position of the modulated bond allows the control of the connection or disconnection of a lattice's section. The switch can be located in an arbitrary bond position of the lattice. 

\begin{figure}[t]
\includegraphics[width=\columnwidth]{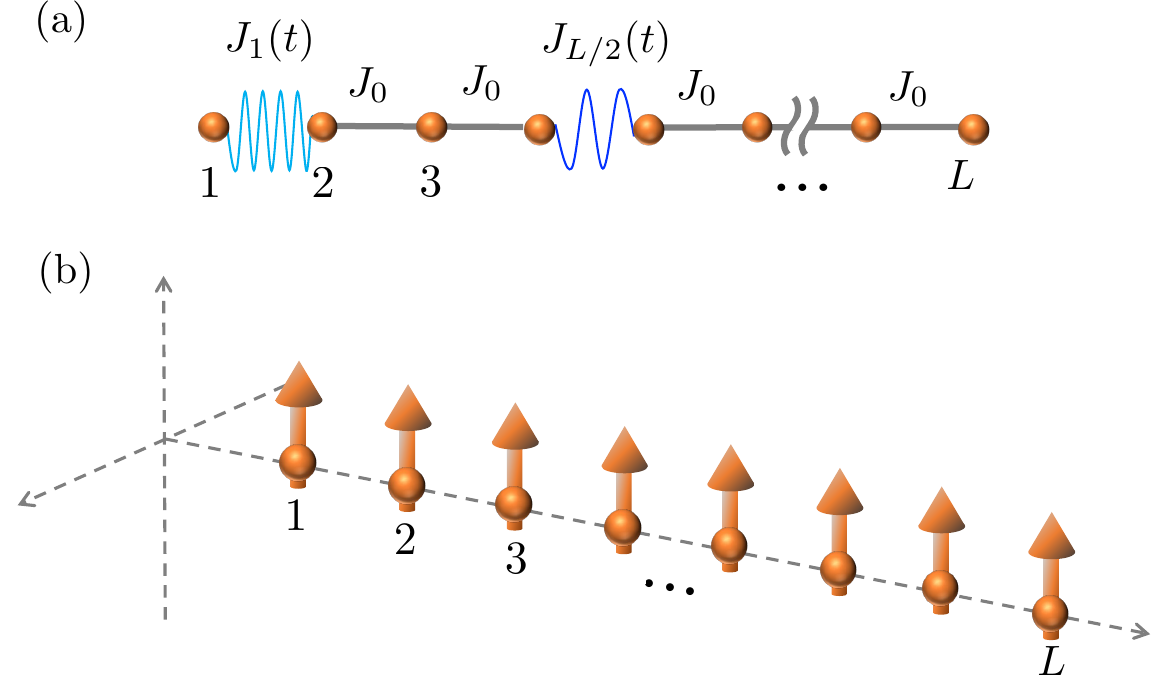}
\caption{(a) The first and $L/2$-th bonds are modulated by $J_{1}(t)=J_{0}\cos{(\Omega_{1}\,t)}$ and $J_{L/2}(t)=J_{0}\cos{(\Omega_{L/2}\,t)}$, respectively, where $L$ is an even integer, $\Omega_{1}=4\,g$, and $\Omega_{L/2}=\Omega_{1}/2$. Light blue and blue wavy lines represent these modulated bonds. The remaining lattice bonds are given by constant spin-spin exchange of  $J_{j}=J_{0}$. (b) Schematic view of the system configuration at $t=0$. The initial condition is the product state $\ket{\psi_0}=\bigotimes_{j=1}^{L}\ket{\uparrow_j}$.
} 
\label{Fig9}
\end{figure}

\begin{figure}
\includegraphics[width=\columnwidth]{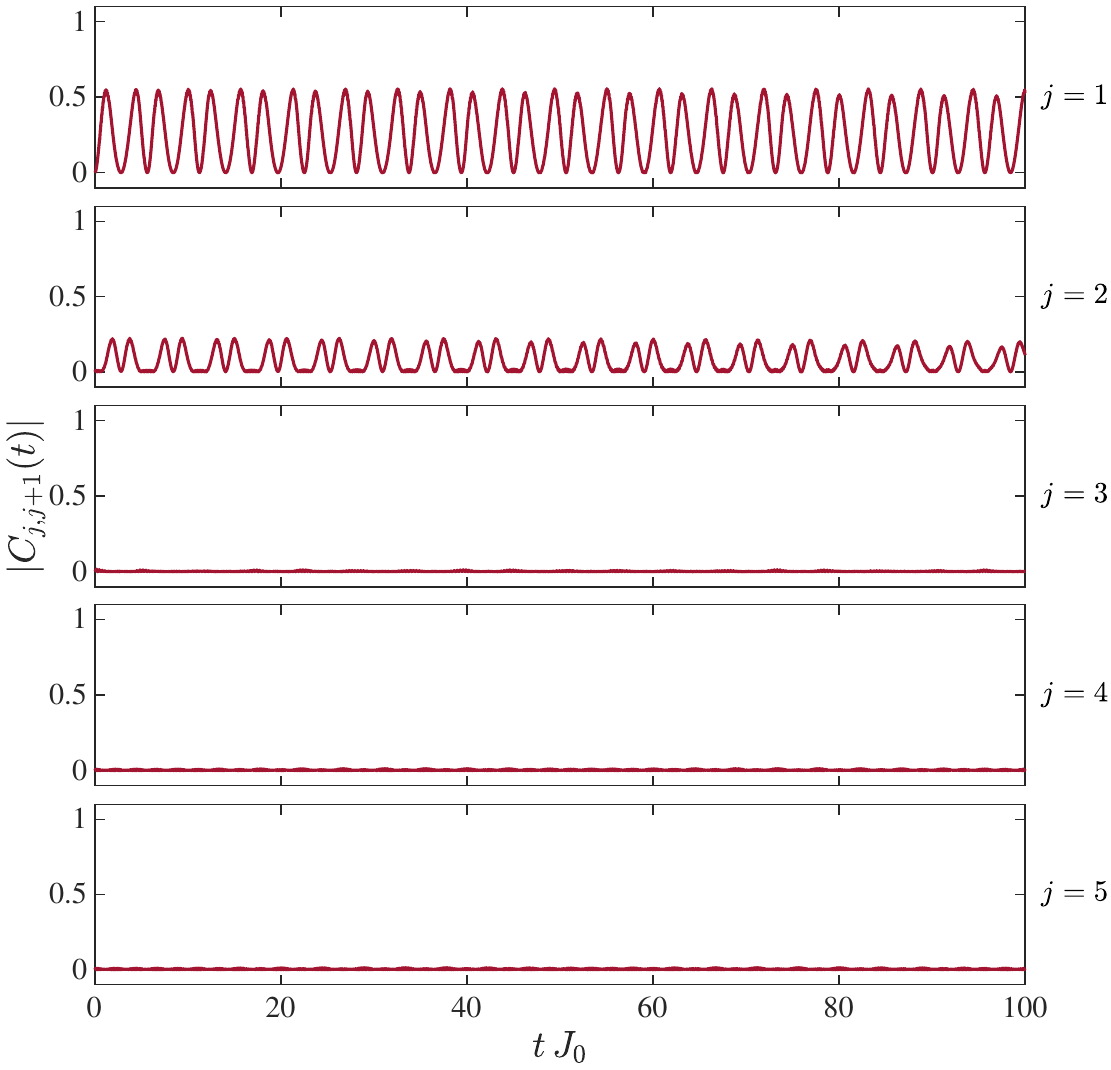}
\caption{The absolute value of correlation function $|C_{j,j+1}(t)|$ as a function of time, for switch off with $\Omega_{L/2}=2\,g$. The panels correspond to correlations function from $|C_{1,2}(t)|$ to $|C_{5,6}(t)|$. The lattice configuration and initial condition are shown in Figs.~\ref{Fig9}(a) and~\ref{Fig9}(b), respectively. In this simulation, we set $L=6$, $J_{0}=0.1\,g$, and $\Omega_{1}=4\,g$.} 
\label{Fig10}
\end{figure}

We explore a second configuration shown in Fig.~\ref{Fig9}(a), where the first modulated bond is $J_{1}(t)=J_{0}\cos(\Omega_1 t)$ and the $L/2$-th bond is $J_{L/2}(t)=J_{0}\cos(\Omega_{L/2} t)$, where $\Omega_{L/2}=2g$ lies within the switch off region (c.f. Fig.\ref{Fig7}). Initially, all spins point up with initial state $\ket{\psi_0}=\bigotimes_{j=1}^{L}\ket{\uparrow_j}$. Here, the frequency $\Omega_{L/2}$ is the switch control parameter, and  $\Omega_1$ allows the creation of an excitation in the lattice. The position of $J_{1}(t)$ allows the spin wave propagation to occur initially from left to right in the lattice. Figure~\ref{Fig10} shows the absolute value of the correlation function $|C_{j,j+1}(t)|$
as a function of time for a lattice with $L=6$ sites. Each panel presents nearest-neighbor correlation pairs. In this case, the spin propagation wave is blocked because the left side of the lattice decouples from the right side, as shown in the panels $j\geq 3$, where $|C_{j,j+1}(t)|$ drops nearly to zero. The last configuration gives an analogous switch behavior with respect to the configuration of Fig.~\ref{Fig6}.

\section{Stroboscopic dynamics}

\begin{figure}
\includegraphics[width=\columnwidth]{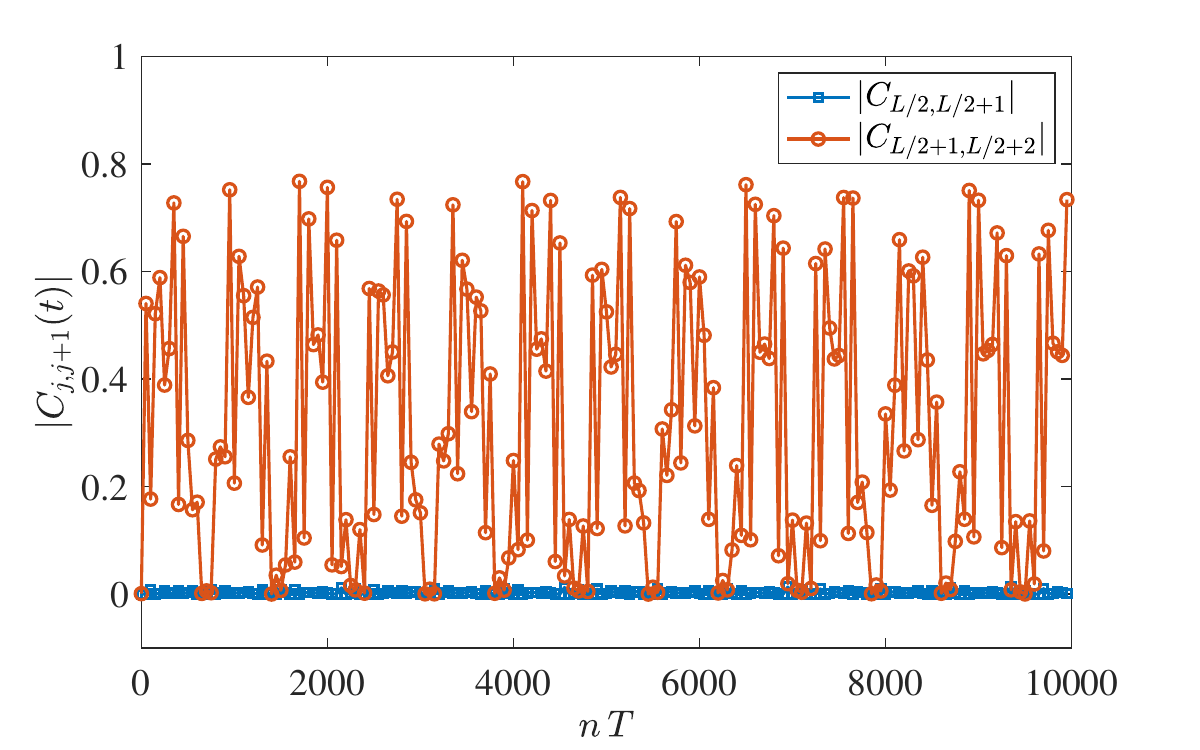}
\caption{Long-time dynamics: The absolute value of the correlation functions $|C_{L/2,L/2+1}(t)|$ (blue square) and $|C_{L/2+1,L/2+2}(t)|$ (red circles) at stroboscopic times $t=n\,T$. The one-period evolution operator in the simulation is calculated with step $\delta t=1.25\times 10^{-5}\,T$ where $T=2\pi/\Omega$. The stroboscopic time vector $t=n\,T$ $\in\,[1.0,10^{4}]$. For clarity, we plot points using the step $\delta t_n= 50\,T$. The parameters are defined in terms of the transverse field $g$, that is, $J_{0}=0.1\,g$, $\Omega=2\,g$, and $L=6$ spins.}
\label{Fig11}
\end{figure}

Here, we expose the long-time behavior of correlation functions (Eq.~\ref{correlation function}) using the stroboscopic dynamics, where the wavefunction reads $\ket{\psi(nT)}=[\hat{U}(T)]^n\ket{\psi_0}$ with $\hat{U}(T)$ the one-period evolution operator \cite{Grifoni1998Oct,Bukov2015Mar}. The latter is numerically computed using exact diagonalization with the configuration shown in Fig.~\ref{Fig6}(a) and the initial condition of Fig.~\ref{Fig6}(b).

The stroboscopic dynamics show that the operational modes of the quantum switch remain in the long term. 
Figure~\ref{Fig11} shows the absolute value of the correlation function at the $L/2$-th bond and the $L/2+1$-th bond at stroboscopic times $t=n\,T$, with a driving frequency of  $\Omega=2\,g$ operating in the switch-off mode. In Fig.~\ref{Fig11}, the correlation $|C_{L/2,L/2+1}(t)|$ remains nearly zero up to $t=10^4\,T$, demonstrating that the switch-off mode is robust in the long term. For clarity in the presentation, the stroboscopic evolution is shown each $\delta t_n=50T$.
To explain the stroboscopic dynamics obtained in our numerical simulations, we calculate the Floquet Hamiltonian for an $L$-spin-lattice using the Magnus expansion \cite{Bukov2015Mar}, which is justified by the high-frequency regime $\Omega\gg J_0$. First, considering the Hamiltonian ($\ref{H1}$), we move to a rotating frame to compute the $L$-spin-lattice Hamiltonian in the interaction picture as follows
\begin{eqnarray}
\hat{H}_{I}(t)=\hbar\sum_{j=1}^{L-1}&&\,J_{j}(t)(\,e^{4igt}\hat{\sigma}_{j}^{+}\,\hat{\sigma}_{j+1}^{+} + \hat{\sigma}_{j}^{+}\,\hat{\sigma}_{j+1}^{-}\\
&&+\,\hat{\sigma}_{j}^{-}\,\hat{\sigma}_{j+1}^{+}+e^{-4igt}\hat{\sigma}_{j}^{-}\,\hat{\sigma}_{j+1}^{-}\,).\nonumber
\label{IntPict}
\end{eqnarray}
The one-period evolution operator reads
\begin{equation}
\hat{U}(T)=e^{-i\hat{H}_{F}T/\hbar}\,,
\end{equation}
where $\hat{H}_{F}$ is the time-independent Floquet Hamiltonian. In the high-frequency regime, $\hat{H}_{F}$ can be approximated using the Magnus expansion \cite{Bukov2015Mar} as $\hat{H}_{F}=\sum_{l=0}^{\infty}\hat{H}^{(l)}_{F}$. The first two terms of the expansion read
\begin{subequations}
\begin{equation}
    \hat{H}_{F}^{(0)}=\frac{1}{T} \int_{0}^{T} dt\,\hat{H}_{I}(t)\,,   
\end{equation}
\begin{equation}
    \hat{H}_{F}^{(1)}=\frac{1}{2!\,i\hbar\,T} \int_{0}^{T}\,dt_{1}\int_{0}^{t_{1}}dt_{2} \,[\hat{H}_{I}(t_{1}),\hat{H}_{I}(t_{2})].
\end{equation}
\label{1Magnus}
\end{subequations}

\begin{figure}[t]
\includegraphics[width=\columnwidth]{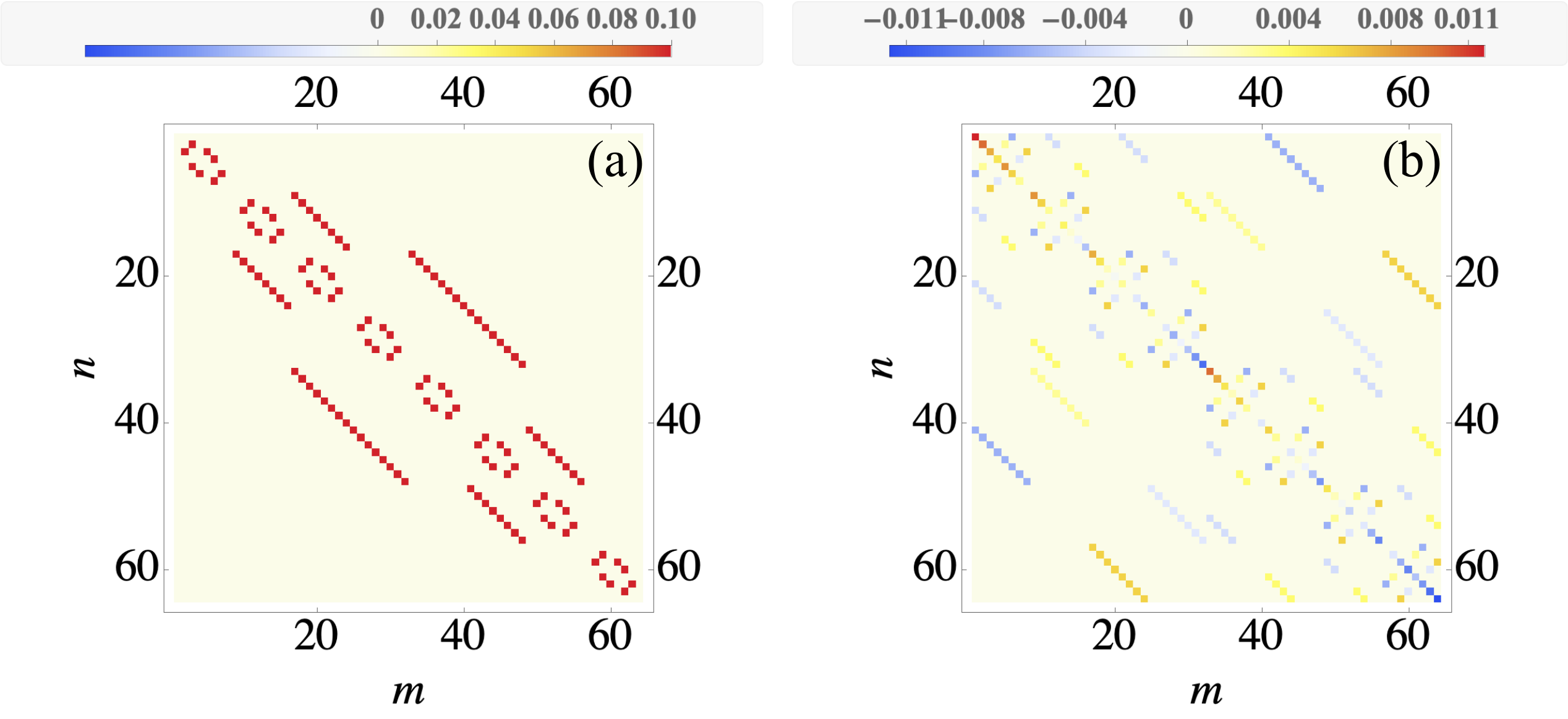}
\caption{Visual representation of the first two components of the Magnus expansion. (a) Represents the matrix elements $(H_F^{(0)})_{n,m}$ whereas (b) stands the matrix elements $(H_F^{(1)})_{n,m}$. The parameters are defined in terms of the transverse field $g$, that is, $J_{0}=0.1\,g$, $\Omega=2\,g$, and we consider a lattice of $L=6$ spins.} 
\label{Fig12}
\end{figure}

Figure \ref{Fig12} shows a visual representation of $\hat{H}_{F}^{(0)}$ and $\hat{H}_{F}^{(1)}$, which have been computed numerically following Eqs.~(\ref{1Magnus}). Here one clearly see that $\hat{H}_{F}^{(0)}$ dominates over $\hat{H}_{F}^{(1)}$ being one order of magnitude larger. Notice that the matrix elements are given in terms of the transverse field $g$. In other words, first-order processes dominate the long-time stroboscopic dynamics with respect to second-order processes, thus providing a robust performance of the quantum switch within the simulated time; see Fig.~\ref{Fig11}. In fact, for the configuration shown in Fig.~\ref{Fig6}, the effective Hamiltonian $\hat{H}_{F}^{(0)}$ can be analytically computed as
\begin{eqnarray}
\hat{H}_F^{(0)}&=&\hbar J_0\sum_{j=1}^{L/2-1}(\hat{\sigma}_{j}^{+}\,\hat{\sigma}_{j+1}^{-}+\hat{\sigma}_{j}^{-}\,\hat{\sigma}_{j+1}^{+})\nonumber\\
&+& \hbar J_0\sum_{j=L/2+1}^{L-1}(\hat{\sigma}_{j}^{+}\,\hat{\sigma}_{j+1}^{-}+\hat{\sigma}_{j}^{-}\,\hat{\sigma}_{j+1}^{+}),
\label{Magnus0}
\end{eqnarray}
which represents two uncoupled $XY$ spin-$1/2$ lattices. 

However, beyond the timescales shown in Fig.~\ref{Fig11}, higher-order corrections in the Floquet expansion are expected to gradually reintroduce effective couplings between the subsystems, ultimately leading to a breakdown of the blocking regime.

In circuit QED architectures, see Ref.~\cite{Zha2020PRL}, transmons exhibit relaxation time $T_1\simeq 74\,\mu$s and dephasing time $T^{*}_2\simeq\,2\mu$s. Also, fluxonium devices exhibit better times $T_1 \simeq 560\,\mu \text{s}$ and $T^{*}_2\simeq 160\,\mu\text{s}$, see Ref.~\cite{Ding2023}. Considering qubit gaps $\omega_{01}=2\pi\times 4\,\text{GHz}$ for a transmon and $\omega_{01}=2\pi\times 0.3\,\text{GHz}$ for a fluxonium, the corresponding oscillation periods ($2\pi/\omega_{01}$) are approximately $T\sim 0.25 \,\text{ns}$ for the transmon and $T\sim 3\,\text{ns}$ for fluxonium. The time window considered in Fig.~\ref{Fig11} spans $10^4\,T$, which corresponds to roughly $10^4\,T = 2.5\,\mu\text{s}$ for transmon and $10^4\,T = 30\,\mu\text{s}$ for fluxonium. These durations are well within the coherence times of their respective platforms, ensuring that the quantum coherent dynamics relevant to our proposal can be observed before decoherence effects become significant. Therefore, the operational timescales required for implementing our switching protocol are compatible with currently achievable coherence times in state-of-the-art superconducting circuits.

\section{Local driving on a single spin}
The fundamental mechanism underlying control via local periodic fields was first reported by Dunlap and Kenkre \cite{Dunlap1986,Dunlap1988Apr}, who demonstrated that in the high-frequency limit, hoppings between neighboring sites can be completely suppressed, a phenomenon known as dynamical localization. This mechanism was subsequently analyzed for more complex quantum lattice systems in Ref.~\cite{Bukov2015Mar}.
Nowadays, state-of-the-art superconducting quantum simulators allow the precise control of physical parameters such as the superconducting gap to engineer nonequilibrium situations without static analog, thus providing an ideal scenario to apply dynamical localization phenomena. 
In Ref.~\cite{Zhao2022Oct}, it has been proven operator spreading via Floquet engineering the superconducting gap of transmon qubits. This Floquet control uses coherent destruction of tunneling \cite{Grossmann1991Jul,Kayanuma2008Jan} of nearest-neighbor qubit-qubit interactions governed by the $XY$ Hamiltonian.
Using the same Floquet control, a recent experimental contribution implemented a synthetic magnetic vector potential in 2D \cite{Rosen2024Dec}. In the time-dependent transverse field Ising model, we discuss the quantum switch using Floquet control on a single spin. This is achieved considering the Hamiltonian
\begin{equation}
H(t)= \hbar \,\sum_{j=1}^{L}g_{j}(t)\,\hat{\sigma}_{j}^{z}+\hbar\,\lambda_{0}\sum_{j=1}^{L-1}\,\hat{\sigma}_{j}^{x}\hat{\sigma}_{j+1}^{x},
\label{HLocal}
\end{equation}
where the spin local driving is 
\begin{align}
g_{j}(t)&=\ \begin{cases}
g+\varepsilon\cos(\nu\,t)\, ,\quad & \text{if}\,\quad j=k\ ,\\
g\ ,\quad & \text{if}\,\quad  j\neq k\,.
\end{cases}
\label{g_local}
\end{align}
\begin{figure}[t]
\includegraphics[width=\columnwidth]{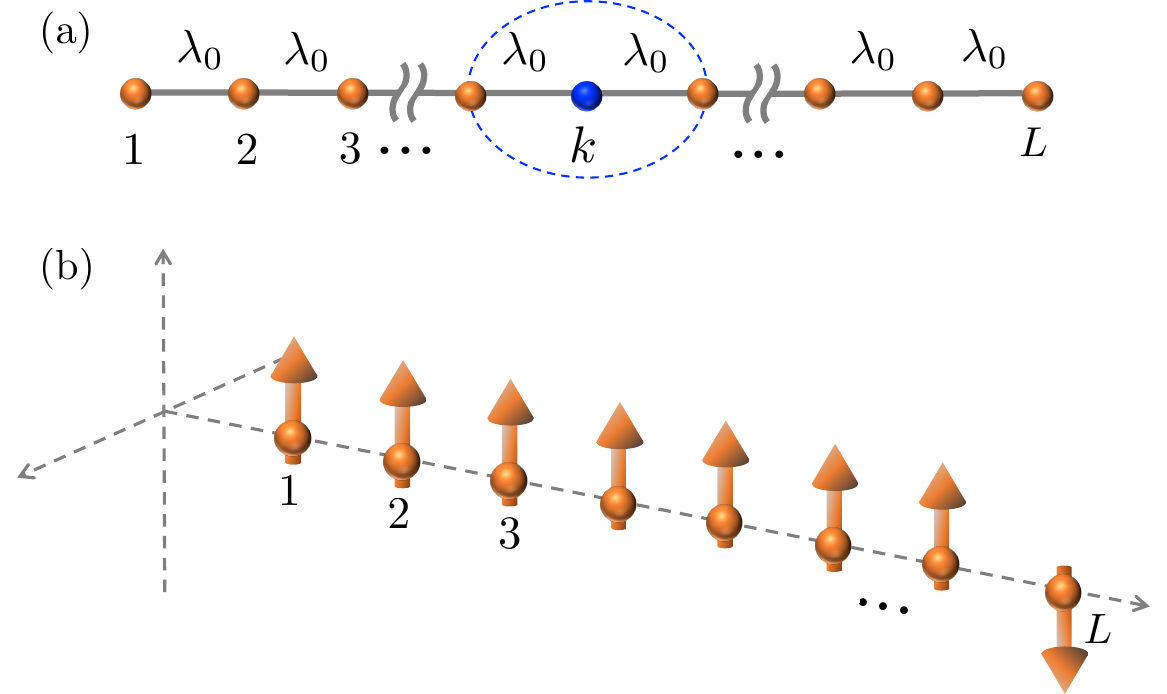}
\caption{(a) Configuration where the spin at position $k$ is driven locally by $g(t)_{L/2}=g+\varepsilon\cos(\nu\,t)$, which is indicated with a blue sphere. The remaining spins have $g_{j}=g$. The lattice bonds are uniform, $\lambda_{0}$. In the switch-off mode, the blue dashed line indicates the $k$th lattice site decouples to the right and left lattice sides. (b) Schematic representation of the system at $t=0$, where the initial state is given by $\ket{\psi_0}=\bigotimes_{j=1}^{L-1}\ket{\uparrow_j}\otimes\ket{\downarrow_L}$.}
\label{Fig13}
\end{figure}
Let us consider the configuration shown in Fig.~\ref{Fig13}(a), where the spin at position $k$ experiences a local driving field, while all other spins have a constant field, $g_{j}=g$. Additionally, all the lattice bonds are assumed constant, denoted by $\lambda_{0}$. 

We move to a rotating frame considering the Hamiltonian Eq.~(\ref{HLocal}), taking into account the local time dependence of the driving at the $k$th spin in Eq.~(\ref{g_local}). The Hamiltonian in the rotating frame reads
\begin{eqnarray}
&\hat{H}_{I}&(t)=\hbar\lambda_{0}\,\sum_{j=1}^{k-2}\,(\,e^{4igt}\hat{\sigma}_{j}^{+}\,\hat{\sigma}_{j+1}^{+} + \hat{\sigma}_{j}^{+}\,\hat{\sigma}_{j+1}^{-}+\text{H.c}.)\nonumber\\\nonumber
&+&\,\hbar\lambda_{0}\,(\,e^{4igt}\,e^{2iF(t)}\hat{\sigma}_{k-1}^{+}\,\hat{\sigma}_{k}^{+} + \,e^{-2iF(t)}\hat{\sigma}_{k-1}^{+}\,\hat{\sigma}_{k}^{-}+\text{H.c}.)\\\nonumber
&+&\,\hbar\lambda_{0}\,(\,e^{4igt}\,e^{2iF(t)}\hat{\sigma}_{k}^{+}\,\hat{\sigma}_{k+1}^{+} + \,e^{2iF(t)}\hat{\sigma}_{k}^{+}\,\hat{\sigma}_{k+1}^{-}+\text{H.c}.)\\
&+&\hbar\lambda_{0}\,\sum_{j=k+1}^{L-1}\,(\,e^{4igt}\hat{\sigma}_{j}^{+}\,\hat{\sigma}_{j+1}^{+} + \hat{\sigma}_{j}^{+}\,\hat{\sigma}_{j+1}^{-}+\text{H.c}.)\,,
\label{IntPictLoc}
\end{eqnarray}
where $F(t)=(\varepsilon/\nu)\sin{(\nu\,t)}$. Using the Jacobi-Anger expansion, $H_{I}(t)$ reads
\begin{eqnarray}\label{HInt_1}
\hat{H}_{I}(t)&=&\hbar\lambda_{0}\,\sum_{j=1}^{k-2}\,(\,e^{4igt}\hat{\sigma}_{j}^{+}\,\hat{\sigma}_{j+1}^{+} + \hat{\sigma}_{j}^{+}\,\hat{\sigma}_{j+1}^{-}+\text{H.c}.)\\\nonumber
&+&\,\hbar\lambda_{0}\,[\,e^{4igt}\,\sum_{m\in\mathbb{Z}}\mathcal{J}_{m}\left(\frac{2\varepsilon}{\nu}\right)e^{im\nu t}\,\hat{\sigma}_{k-1}^{+}\,\hat{\sigma}_{k}^{+} \\\nonumber
&+&\sum_{m\in\mathbb{Z}}\mathcal{J}_{m}\left(\frac{2\varepsilon}{\nu}\right)e^{-im\nu t}\,\hat{\sigma}_{k-1}^{+}\,\hat{\sigma}_{k}^{-}+\text{H.c}.]\\\nonumber
&+&\,\hbar\lambda_{0}\,[\,e^{4igt}\,\sum_{m\in\mathbb{Z}}\mathcal{J}_{m}\left(\frac{2\varepsilon}{\nu}\right)e^{im\nu t}\,\hat{\sigma}_{k}^{+}\,\hat{\sigma}_{k+1}^{+} \\\nonumber
&+&\sum_{m\in\mathbb{Z}}\mathcal{J}_{m}\left(\frac{2\varepsilon}{\nu}\right)e^{im\nu t}\,\hat{\sigma}_{k}^{+}\,\hat{\sigma}_{k+1}^{-}+\text{H.c}.]\\\nonumber
&+&\hbar\lambda_{0}\,\sum_{j=k+1}^{L-1}\,(\,e^{4igt}\hat{\sigma}_{j}^{+}\,\hat{\sigma}_{j+1}^{+} + \hat{\sigma}_{j}^{+}\,\hat{\sigma}_{j+1}^{-}+\text{H.c}.)\,,
\end{eqnarray}
with $\mathcal{J}_{m}\left(2\varepsilon/\nu\right)$ the $m$-th Bessel function of the first kind.  
\begin{figure}[t]
\includegraphics[width=\columnwidth]{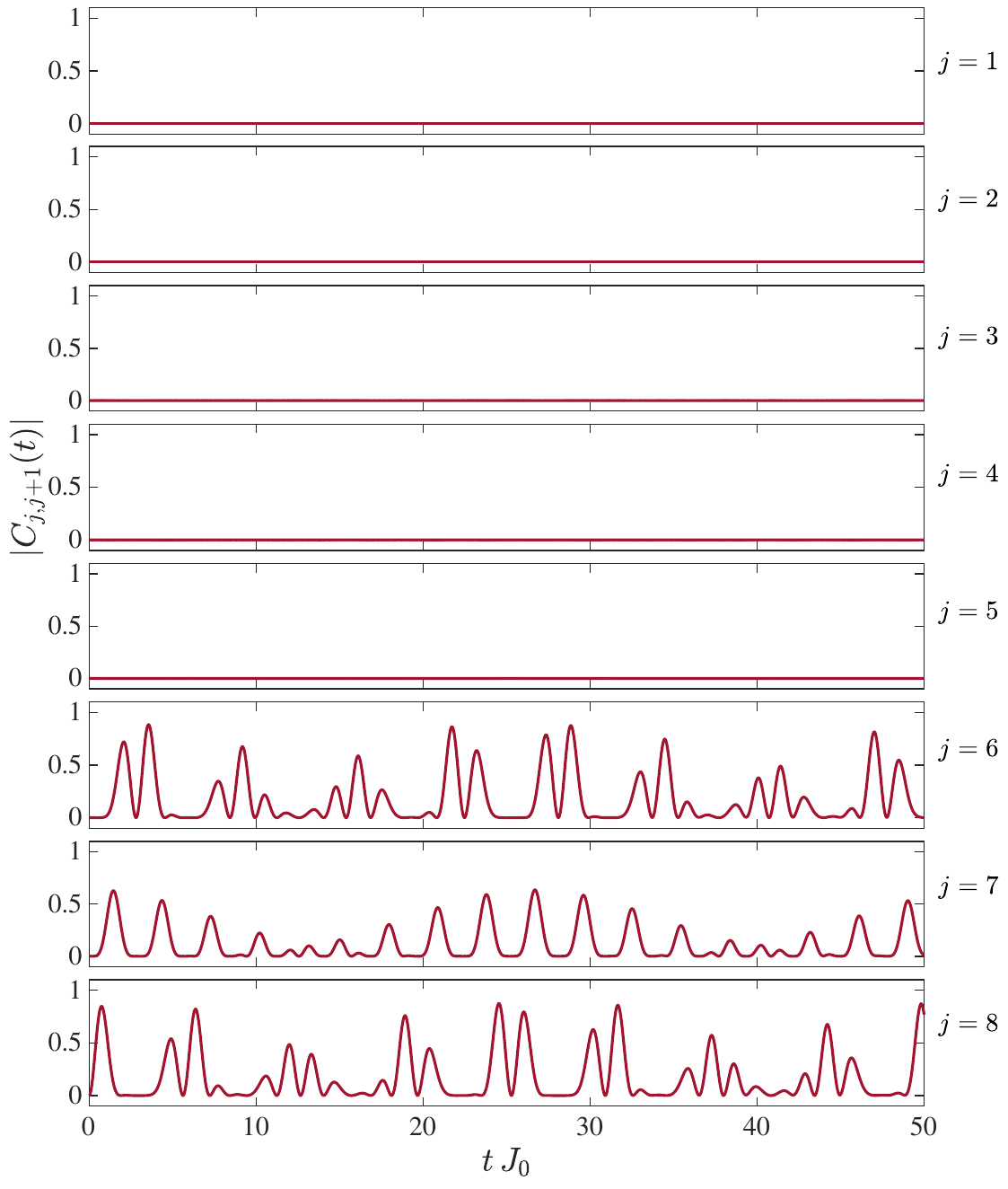}
\caption{Switch-off mode using local driving on a single spin: The absolute value of correlation function $|C_{j,j+1}(t)|$ as a function of time. The panels correspond to correlations function from $|C_{1,2}(t)|$ to $|C_{8,9}(t)|$. The lattice configuration and initial condition are shown in Figs.~\ref{Fig13}(a) and~\ref{Fig13}(b). The parameters of the local driving are $k=5$, $\nu=3\,g$ and $\varepsilon= (x_{0}\,\nu)/2$, where $x_{0}=2.4048$ is the first root zeroth Bessel function. In this simulation we set $L=9$, $\lambda_{0}=0.01\,g$, and $\delta t=0.005\lambda_0^{-1}$.}
\label{Fig14}
\end{figure}
We select the control parameters within the rotating wave approximation (RWA) to implement the quantum switch using a single locally driven spin. This approach simplifies the Hamiltonian in Eq.(\ref{HInt_1}) into two uncoupled XY spin-1/2 lattices, effectively allowing us to neglect the rapidly oscillating terms. First, we operate in the frequency regime where $g\gg\lambda_{0}$, which ensures that terms proportional to $e^{\pm 4igt}$ oscillate faster than processes governed by the spin-spin interaction $\lambda_0$ and can be neglected. Second, by choosing $\nu\gg\lambda_{0}$ and ensuring that $\nu=m g$ with $m\ge 3$ is an odd integer, terms involving $e^{\pm 4igt}\,e^{\pm im\nu t}$ are negligible under the RWA. Finally, only the terms proportional to $e^{\pm im\nu t}$ where $m=0$ become resonant, leading to the following effective Hamiltonian in the rotating frame
\begin{eqnarray}\label{HInt_2}
\hat{H}_{I}(t)&=&\hbar\lambda_{0}\sum_{j=1}^{k-1}(\hat{\sigma}_{j}^{+}\,\hat{\sigma}_{j+1}^{-}+\hat{\sigma}_{j}^{-}\,\hat{\sigma}_{j+1}^{+})\\\nonumber
&+& \hbar \lambda_{0}\mathcal{J}_{0}\left(\frac{2\varepsilon}{\nu}\right)(\hat{\sigma}_{k-1}^{+}\,\hat{\sigma}_{k}^{-}+\hat{\sigma}_{k-1}^{-}\,\hat{\sigma}_{k}^{+})\\\nonumber
&+& \hbar \lambda_{0}\mathcal{J}_{0}\left(\frac{2\varepsilon}{\nu}\right)(\hat{\sigma}_{k}^{+}\,\hat{\sigma}_{k+1}^{-}+\hat{\sigma}_{k}^{-}\,\hat{\sigma}_{k+1}^{+})\\\nonumber
&+&\hbar\lambda_{0}\sum_{j=k+1}^{L-1}(\hat{\sigma}_{j}^{+}\,\hat{\sigma}_{j+1}^{-}+\hat{\sigma}_{j}^{-}\,\hat{\sigma}_{j+1}^{+}).\nonumber
\end{eqnarray}
The Hamiltonian in Eq.(\ref{HInt_2}) represents an equivalent system of two XY spin-1/2 lattices which are coupled to the $k$th spin with renormalized spin-spin coupling $\lambda_0\mathcal{J}_{0}\left(2\varepsilon/\nu\right)$. This way, by choosing $\varepsilon= (x_{0}\,\nu)/2$, where $x_{0}=2.4048$ is the first root zeroth Bessel function, the effective Hamiltonian represents two uncoupled XY spin-1/2 lattices provided by the coherent destruction of tunneling in a quantum many-body system \cite{Grossmann1991Jul,Kayanuma2008Jan,Gong2009Sep}. 

Uncoupling the XY spin-1/2 lattices using the local spin driving allows the system to operate as a switch-on and off. To see this, we numerically solve the Schr\"odinger  equation using the Hamiltonian (\ref{HLocal}) for $L=9$ spins initialized in the state $\ket{\psi_0}=\bigotimes_{j=1}^{L-1}\ket{\uparrow_j}\otimes\ket{\downarrow_L}$ and a local driving acting upon the spin at position $k=5$. The parameters for this simulation are specified in the caption of Fig.~\ref{Fig14}, whose panels show the nearest-neighbor correlation function for each lattice site $j$. The spin wave originates on the right side and propagates from right to left. The local driving causes the right side of the lattice to become uncoupled from the left, resulting in correlation functions approximately equal to zero for $j<6$. Controlling a single spin causes the correlation function between the spin at site $k$ and its two nearest neighbors to drop to zero. In this way, controlling a single spin allows an analogous switch behavior, as is the case for modulated bonds previously exposed.

Remarkably, we identify an integration between the local driving and the modulated bonds with zero time average discussed in Section~\ref{ModulBond}. A local driving acting upon a single spin (qubit) will effectively renormalize the spin-spin interaction via Bessel functions, that is, $\lambda_0 \sum_{m\in \mathbf{Z}}\mathcal{J}_m(2\varepsilon/\nu)e^{\pm i m\nu t}$, see Eq.~(\ref{HInt_1}). As is well established in the literature \cite{Bukov2015Mar,Dunlap1986,Dunlap1988Apr}, if the local driving frequency $\nu\gg \lambda_0$, we can invoke the rotating-wave approximation so that the strength of the spin-spin interaction becomes $\lambda_0\mathcal{J}_0(2\varepsilon/\nu)$. So, suppose the Bessel function oscillates around its first zero ($x_0=2.4048$) at a given frequency $\omega$. In that case, one may realize an atypical spin-spin interaction with an average zero within $T=2\pi/\omega$. This idea was first exposed in Ref.\,\cite{Pieplow2018Jul}; however, no specific control function was provided. Here, we propose the following control function

\begin{align}
    \frac{2\varepsilon(t)}{\nu} = x_1\cos(\omega t) + \bigg(\frac{x_1+x_2}{2}\bigg)[1-\cos(\omega t)]\coloneqq F(t,\omega),
\end{align}
where $x_1=2$ and $x_2=2.84787695$.
 
\begin{figure}
\includegraphics[width=1.\columnwidth]{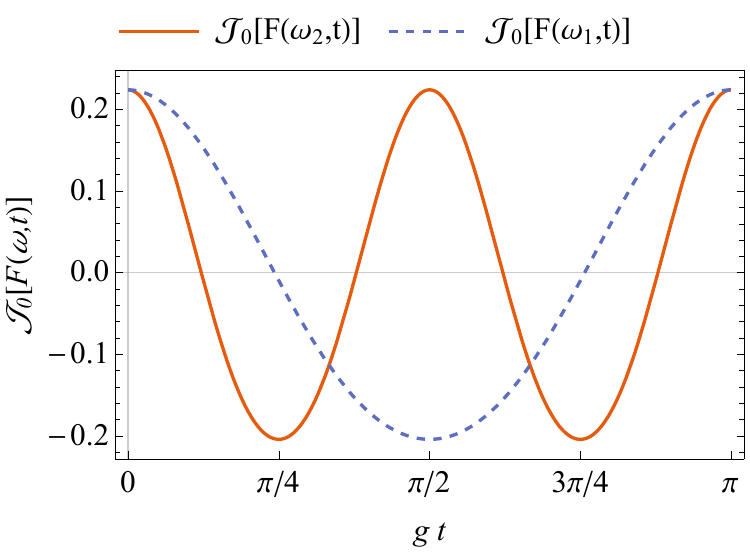}
\caption{Periodic behavior of the Bessel function $\mathcal{J}_0(F(t,\omega))$ for parametric resonances $\omega_1=2g$ and $\omega_2=4g$. Both control functions $F(\omega_2,t)$ and $F(\omega_1,t)$ are periodic with period $T=\pi/g$, which corresponds to the discrete periodicity of the Hamiltonian that describes the situation shown in Fig.~\ref{Fig1}.}
\label{Fig15}
\end{figure}

In our study, we want the spin-spin coupling strength to oscillate at the resonance frequencies $\omega_1=2g$ and $\omega_2=4g$, as we proposed in Fig.~\Ref{Fig1}. Figure~\ref{Fig15} shows the behavior of the Bessel function for control functions $F(t,\omega_1)$ and $F(t,\omega_2)$. We see an oscillating Bessel function that implements the desired modulated bonds. Also, it can be shown using Mathematica that the time average of the Bessel function in one period ($T=\pi/g$) is of the order of $10^{-8}$ for both oscillating frequencies $\omega_1$ and $\omega_2$.       

\section{Conclusion}
In summary, we implemented coupling and decoupling mechanisms for system and reservoir interaction using Floquet control in the time-dependent transverse field Ising model. We analyzed scenarios where the driving frequencies act either modulating bonds between spins or a single spin. Different driving configurations and initial conditions were explored, focusing on the blocking and deblocking effects in the propagation of spin and correlation waves.

Moreover, we implemented the coupling and decoupling mechanism using a single-modulated bond. The operational modes of the lattice were characterized over a wide frequency range, identifying specific frequencies where the device toggles between on and off modes. The stroboscopic dynamics of correlation functions demonstrated the long-term robustness of the quantum switch's operational modes.

Finally, we realized the coupling-decoupling mechanism using Floquet control applied to a single spin. This approach allowed us to control the coherent destruction of tunneling, effectively decoupling the spin-lattice into two independent XY spin- 1/2 lattices. 

Our analysis is conducted in a high-frequency regime, where hopping oscillates at frequencies $\Omega\gg J_{0}$ with $J_{0}=0.1g$ being the bare hopping rate, and the rotating-wave approximation becomes applicable. However, it is noteworthy that a realistic circuit QED implementation makes the infinite-frequency limit impractical due to inherent experimental constraints. For instance, accessible driving frequencies for on-chip spectroscopy are on the order of $2\pi\, \times 20 \,\text{GHz}$ \cite{DiCarlo2009,DiCarlo2010,Reed2012}. At the same time, the qubit gap $g$ is tunable in the range $g\in [2\pi \times 4- 2\pi \times 4.6] \,\text{GHz}$ for transmons \cite{Zha2020PRL}, or $g\in [2\pi \times 0.3- 2\pi \times 1.0]\,\text{GHz}$ for fluxonium devices \cite{Ding2023, Bao2022}. In our simulations, e.g., Fig.~\ref{Fig7} and Fig.~\ref{Fig14}, we explore driving frequencies up to $\Omega=3g$ and $\nu=3g$, respectively, which lies well within the upper bound of what can be realized experimentally.
Crucially, within the frequency ranges mentioned above, our model enables access to the system's parametric resonances at $\Omega_0=2g$ and $\Omega_1=4g$, a regime in which nontrivial dynamical behavior arises. Such phenomena are inherently beyond the scope of treatments restricted to the infinite-frequency limit.

Our proposed schemes exhibit analogous switch behavior, exhibiting versatility and robustness. Our results highlight the potential for implementing coupling-decoupling mechanisms in state-of-the-art quantum technologies with applications across various current research fields.


\begin{acknowledgements}
We thank F. Albarrán-Arriagada for his constant support. We also thank A. Bayat and V. Montenegro for useful discussions. M.~A. acknowledges financial support from ANID Postdoctoral FONDECYT Grant No. 3240443. In addition to Vicerrectoría de Investigación, Innovación y Creación (VRIIC) USACH Grant No.~POSTDOC\_DICYT 042131RA\_POSTDOC. G. Romero acknowledges support from Dicyt USACH under grant $042331{\rm RH\textunderscore Ayudante}$.
\end{acknowledgements}

\bibliographystyle{apsrev4-1}
\bibliography{references}

\end{document}